\documentclass[12pt]{iopart}

\usepackage[subnum]{cases}
\usepackage{iopams}
\usepackage{graphicx}
\newcommand{\no}{\nonumber}
\newcommand{\pa}{\partial}
\newcommand{\ka}{\kappa}
\newcommand{\la}{\lambda}
\newcommand{\ra}{\rightarrow}

\newcommand{\fxdpt}{
\begin{figure}[t]
\begin{center}
 \includegraphics[width=0.5\textwidth,clip]{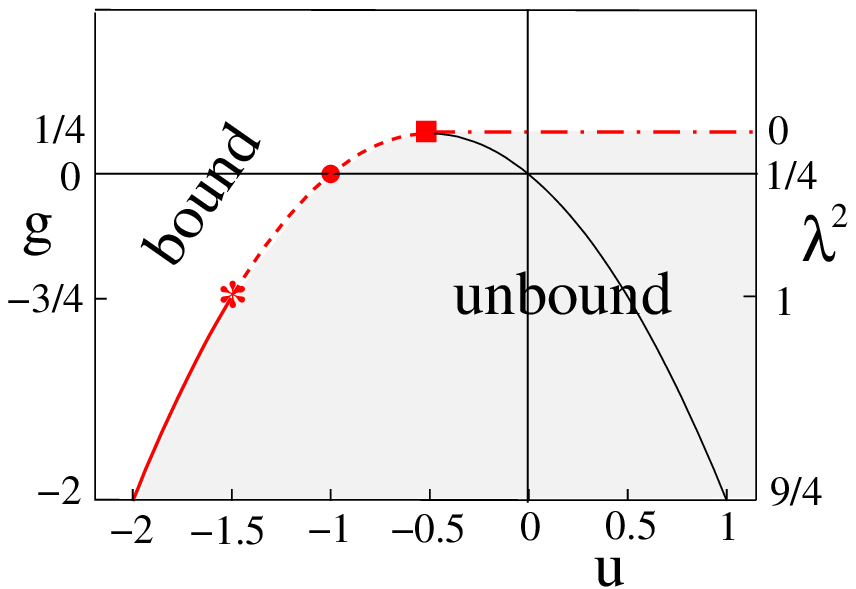}
\end{center}
\caption{$g$ {\it vs.} $u$ phase diagram. The plot shows the phases
  and the RG fixed points in the $g$-$u$ plane ($u=-V_0\,a^2$). The
  red curve below $g=1/4$ and $u=-0.5$ show the binding-unbinding
  transitions governed by a line of unstable real fixed points.  The
  transition is first order for $g<-3/4$ and second order for
  $-3/4<g<1/4$. This line is the transition line in the limit of zero
  range potential ($a\rightarrow 0$, $V_0\rightarrow \infty$, with
  $u=$constant) The black continuous curve for $u>-0.5$ shows the
  locus of stable fixed points representing the unbound phase. The
  dashed-dotted line at $g=1/4$ is the boundary beyond which the fixed
  points are complex.}
\label{fig:fxdpt}
\end{figure}
}
\newcommand{\sla}{
\begin{figure}[hb]
\begin{center}
\includegraphics[width=0.5\textwidth,clip]{sadhukhan_fig2.eps}
\end{center}
\caption{$S$ {\it vs.} $\la$ for various $\ka$. The plot shows that
  the entropy diverges for $\la\le0$ as $\ka\ra 0$. The dashed line
  marked as $\kappa\ra 0$ is the expected behaviour of the entropy for
  $\lambda>1$. }
\label{fig:sla}
\end{figure}
}
\newcommand{\skap}{
\begin{figure}[b]
\begin{center}
\includegraphics[width=0.5\textwidth,clip]{sadhukhan_fig3.eps}
\end{center}
\caption{$S$ {\it vs.}  $\ln\,\ka$ for various $\la$. For
  comparison, $3\ln\ka$ and $\frac{3}{2}\ln\ka$ are shown by black
  lines with the symbols $+$ and $\times$. The inset shows that the
  entropy is $\ka$-independent for $\la>1$.}
\label{fig:skap}
\end{figure}
}
\newcommand{\clam}{
\begin{figure}[ht]
\begin{center}
\includegraphics[width=0.5\textwidth,clip]{sadhukhan_fig4.eps}
\end{center}
\caption{The plot of $c_\la$ {\it vs.} $\la$,
  showing a divergence at $\la=1$.}
\label{fig:clam}
\end{figure}
}
\newcommand{\col}{
\begin{figure}[t]
\begin{center}
 \includegraphics[width=0.5\textwidth,clip]{sadhukhan_fig5.eps}
\end{center}
\caption{Data collapse: $(S_\lambda-S_1)/\frac{3}{2}\ln\ka$ vs
  $(1-\la)\ln\ka$.}
\label{fig:col}
\end{figure}
}
\newcommand{\limit}{
\begin{figure}[t] 
\begin{center}
 \includegraphics[width=0.35\textwidth,clip]{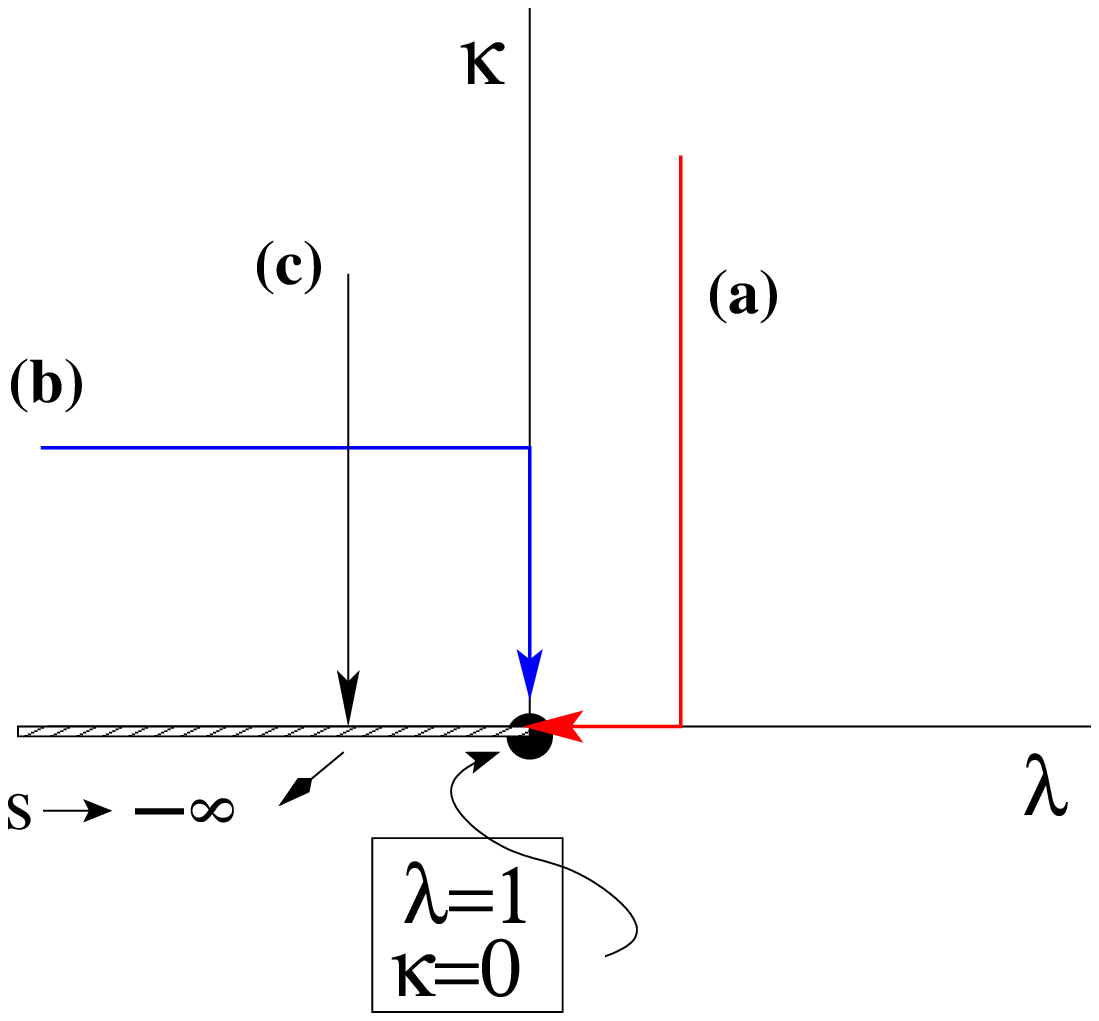}
\end{center}
\caption{The path dependence of entropy. Two different limits of
  approaching $\{\la=1,\ka\ra 0\}$. (a) First $\ka\ra 0$ and then
  $\la=1$ (red line). The entropy diverges like $1/(1-\la)$. (b) First
  $\la\ra 1$ and then $\ka\ra 0$ (blue line). The entropy diverges
  like $\ln\ka$ Thick line along $x$-axis for $\la\le 1$ denotes
  divergent entropy. (c) For $\la<1$, taking the limit $\ka\ra 0$
  (black vertical line with arrow) leads to divergent entropy and $S$
  remains so along the horizontal stretch.}
\label{fig:limit}
\end{figure}
}

\begin{document}
\title[Entanglement entropy for inverse square interaction] {
  Signature of special behaviours of $1/r^2$ interaction in the quantum
  entanglement entropy } 
\author{Poulomi Sadhukhan and Somendra M.  Bhattacharjee}
\ead{poulomi@iopb.res.in, somen@iopb.res.in} 
\address{Institute of Physics, Bhubaneswar, 751 005, India.}

\date{\today}

\begin{abstract}
  We study the bipartite von Neumann entropy of two particles
  interacting via a long-range scale-free potential $V(r)\sim -g/r^2$
  in three dimensions, close to the unbinding transition.  The nature
  of the quantum phase transition changes from critical ($-3/4<g<1/4$)
  to first order ($g<-3/4$) with $g=-3/4$ as a multicritical point.
  Here we show that the entanglement entropy has different behaviours
  for the critical and the first order regimes.  But still there
  exists an interesting multicritical scaling behaviour for all $g\in
  (-2<g<1/4)$ controlled by the $g=-3/4$ case.
\end{abstract}

\submitto{\JPA}
\maketitle

\section{Introduction}

Quantum entanglement\cite{indrani,kitaev,horo,amico} is an important
aspect of quantum mechanics that tells us about the quantum
correlation of two particles or subsystems spatially apart. When a
composite quantum system is in a pure non-product state, then even if
the subsystems are spatially far apart and non-interacting, the
measurement on one subsystem affects that on the other
instantaneously. This ``spooky action at a distance'' later gave birth
to the term ``entanglement''.  This phenomenon was first marked by
Einstein, Podolsky and Rosen in a gedanken experiment\cite{epr}, known
as the EPR paradox. In their paper, they considered two particles
which interacted for some time and showed that it is possible to
measure the conjugate non-commutating quantities, like position and
momentum, simultaneously, in violation of quantum mechanics. Later on,
it was resolved and resulted the idea of the quantum entanglement which
indicates the presence of inherent quantum effects, not just
correlations between the two particles. In fact, quantum entanglement
is a resource for quantum computation with no classical counter-part.
Here, in this paper, we consider two interacting particles in a pure
state, like an EPR pair, or more specifically in the two-particle
ground state, in which case we can expect to observe the entanglement.

A quantum phase transition (QPT) occurs at zero temperature and at the
QPT the ground state energy is non-analytic with respect to some
parameter in the Hamiltonian. A QPT is fully governed by quantum
fluctuations and hence one would expect that the quantum entanglement
would show special signatures at the QPT. Even it is found that the
entanglement entropy behaves in different ways for a first order and a
continuous transition. The critical behaviour of the entanglement
entropy is drawing much attention now-a-days\cite{jphysa} and has been
investigated for different spin models \cite{indrani,kitaev,osterloh}
as well as in continuum systems\cite{deh}.

To quantify the entanglement, various definitions of entanglement
entropy are used, among which the von Neumann
entropy\cite{indrani,kitaev,jphysa,osterloh} is the most common.
Recently it was shown that the von Neumann entropy of two particles
has a $d\ln\ka$ behaviour at the quantum critical point (QCP) of
unbinding in dimensions \hbox{$1<d<4$}. This has been established
analytically for a $3$D potential well\cite{psmb}. Here the QCP is
attained when the inverse length scale $\ka$, the inverse of the width
of the wave function, approaches zero. This is achieved by tuning the
potential or the mass.  Also it is shown that this divergence is
essential for the criticality and linked to the reunion behaviour of
two polymers in the equivalent classical statistical mechanical
problem of polymers. In this paper, we study how the von Neumann
entropy for a long-range potential, like $1/r^2$, changes as one
varies its strength and sign. The equivalent classical statistical
mechanical problem involves two directed polymers interacting at the
same contour length like a DNA with native base pairing but with an
additional $1/r^2$ interaction.  The polymer model has been studied
using renormalization group in Ref.\cite{sm,kol}.

This paper considers two particles interacting through a
three-dimensional inverse square law potential, and finds the quantum
entanglement between the particles. Here we use particle
partitioning\cite{jphysa}.  The Hamiltonian for the two particles we
shall be using is,
\begin{equation}
  H=\frac{{\bf p}_1^2}{2m_1}+\frac{{\bf p}_2^2}{2m_2}+
  V({\bf r}_1-{\bf r}_2),
\end{equation}
where $m_i$, ${\bf r}_i$ and ${\bf p}_i$ are the mass, the position and the
momentum of the $i$th particle,
\begin{equation}{
V(r)=
}\cases{
  -V_0, \quad\quad {\rm for}\ \  r<a,\quad (V_0>0),\\
  -\frac{2\mu}{\hbar^2}\frac{g}{r^2}, \quad {\rm for}\ \ r>a,}
\label{eq:pot}
\end{equation}
is a central potential, and $\mu=m_1m_2/(m_1+m_2)$ is
the reduced mass of the two particles. We take $2\mu/\hbar^2=1$.

The inverse square potential is of immense importance in quantum
mechanics\cite{land}. It is at the boundary of the short and the long
range potentials. For potentials decaying like $r^{-p}$ in three
dimensions, there is no finite bound state if $p>2$, while for a slower
divergence, i.e., $p<2$, there is a finite negative lower bound in
energy. For an attractive potential $-g/r^2$ ($g>0$), the kinetic and
the potential energies are of the same order near small $r$ and so the
bound state spectrum depends on the value of $g$. A manifestation of
the borderline case is in the scale-free nature, $H(\la
r)=\la^{-2}H(r)$.  This makes, $g$, the dimensionless strength of the
potential, a ``marginal'' parameter in the RG sense in all dimensions.
The singularity of $g/r^2$ at the origin prevents discrete bound
states.  A suitable modification of the potential at small
$r$, e.g. by putting a cut off and replacing the potential by a short
range finite attraction near origin, gives discrete bound states. This is
done in Eq.~\ref{eq:pot}.

It is established in quantum mechanics that there is no finite energy
ground state for $g>1/4$. For $g<1/4$, the wave function is
normalizable and the bound state energy can be obtained by the
standard procedure.  In this regime of $g$, the unbinding transition
is induced by tuning the strength of the short-range potential near
$r=0$, depicting the quantum phase transition. The unbinding
transition in this long-range interaction is a unique example of a QPT
whose type can be first order ($g<-3/4$), critical but non-universal
($g>-3/4$), and even Kosterlitz-Thouless type ($g=1/4$)\cite{kol}.
The solvability and the wide repertoire of QPT behaviour make this
model an ideal terrain for exploration of the nature of entanglement
entropy around a QPT\footnote{What makes the unbinding transition
  special is the change in the topology of the underlying space. Note
  that the wave function $|\psi\rangle$ belongs to a space $H_B\ra
  H_c\otimes H_2$ where $H_c$ is the Hilbert space for a free particle
  while $H_2$ is the separable space of square integrable functions.
  This space for the bound state goes over to the space
  $H_U=H_c\otimes H_c$ for two particles (scattering states). Another
  way of seeing this is via currents. The space $H_B$ allows only one
  current for the CM while $H_U$ allows two currents for the two
  unbound particles.  In this paper we stay in the $H_B$ space on the
  bound side.}.  This is what we set to do in this paper.

A phase transition is defined as a singularity in the energy,
associated with diverging length scales. In this sense the quantum
unbinding transition is a genuine phase transition. This QPT exists
because time of infinite extent plays a role in quantum mechanics. It
becomes clear in the path integral formulation. The quantum problem
can be mapped onto an equivalent classical statistical mechanical
problem of polymers under the imaginary time transformation
($it\rightarrow N$). The time in the quantum problem then becomes the
length of the polymer, $N$, Green's function maps on to the partition
function and the ground state energy corresponds to the free energy
per unit length. The interaction between the polymers means the
interaction of a pair of monomers at the same index along the length
of the polymers as in DNA base pairing. This is equivalent to the same
time interaction of two quantum particles.  The equivalent classical
problem in the context of melting transition of two polymers
interacting via a potential like Eq.~\ref{eq:pot} has been discussed
in Ref.~\cite{sm} which reveals that the results of the quantum
problems can be recovered from such studies. Like the quantum particle
making excursion inside and outside of the well, the polymers also
come closer, they reunite, and move further, forming swollen bubbles.
The entropy of a bubble of length $N$ is
\begin{equation}
  \ln\Omega(N)= N\sigma_0-\Psi\ln N,
\end{equation} 
where $\Omega(N)$ is the reunion partition function of two polymers
starting together, reuniting anywhere in space again at length $N$,
$\sigma_0$ is the bubble entropy per unit length and $\Psi$ is the
reunion exponent. The details can be found in
Refs.~\cite{psmb,sm,fish}.

The binding-unbinding transition of polymers, for DNA melting, has been
studied in the context of the necklace model of polymers and it is
found that the reunion exponent $\Psi$ determines the order of
transition\cite{fish}.  The phase transition occurs if $\Psi\ge 1$.
The transition is continuous if $1<\Psi<2$, while it is first order
for $\Psi>2$. In three dimensions, the reunion exponent is given
by\cite{sm}
\begin{equation}
  \Psi=1+\la,\quad {\rm with}\quad \la=\sqrt{\frac{1}{4}-g},
\label{eq:glam}
\end{equation}
where the dependence on $g$, a bit counter-intuitive, is a consequence
of its marginality. Here also we use the parameter $\la$($>0$) because
of its occurrence in the sequel. Table $1$ gives the correspondence
between $g$ and $\la$ for easy reference.
\begin{table}
\begin{center}
\begin{tabular}{|c|c|c|c|c|c|}
\hline
$g$ &-$2$& $-3/4$ &0& $1/4$ &$>1/4$\\
\hline$\lambda$&$1.5$ &$1$&1/2& $0$& imaginary\\
\hline
\end{tabular}
\end{center}
\caption{ Conversion table for $g$-$\la$. Real values of $\la$ occur
  for $g<\le 1/4$.}
\end{table}

\fxdpt

The phase diagram and the lines of RG fixed points are shown in
Fig.~\ref{fig:fxdpt}. This plot shows the phases in the $g$-$u$ plane,
where
\begin{equation}
  u=-V_0\, a^2,
\end{equation} 
in the unit of \hbox{$2\mu /\hbar^2=1$}, is the dimensionless short
range potential in which the two particle state is in. The fixed
points shown here are obtained from the renormalization group analysis
done in Ref.~\cite{sm}. The red line for $u<-0.5$ shows the unstable
fixed points across which the unbinding transition takes place, and
the black continuous line for $u>-0.5$ shows the phases by stable
fixed points. For $g<-3/4$, the bound-unbound transition is first
order for $\Psi>2$, which is indicated by the red continuous line
ending at the symbol * at $g=-3/4$, or, $\la=1$, a multicritical point.
After that, the transition is continuous upto $g=1/4$ with $\Psi<2$.
Beyond that, where $\la$ is imaginary, there is no real fixed point,
and the system is in a bound state.  Across the $g=1/4$ line, with
$u\ge -0.5$, a Kosterlitz-Thouless type phase transition from the
bound to the unbound state can be induced by tuning $g$. The two
regimes, $\Psi<2$ and $\Psi>2$ are governed by different behaviours,
with additional log-corrections at $\Psi=2$.

We find that the entanglement entropy also carries this signature of
the specialilty of $g=-3/4$ or $\la=1$. The entropy in the three
different regimes, $\la<1$, $\la=1$ and $\la>1$, scale in different
manners. We establish that $\la=1$ behaves like a multicritical point
for the entanglement entropy too, controlling both the first order and
the critical behaviour in the whole range $-2\le g\le 1/4$.

The outline of the paper is as follows. In section 2, we describe our
model and the method by which we calculate the von Neumann entropy.
The analytical results are presented in section 3 and the von
Neumann entropy is calculated for $\la<1$. Next we present the exact
numerical results done in {\sc mathematica} and discuss the behaviour
of the entropy and its scaling in section 4. Finally we conclude in
section 5.

\section{Model and method}
Eq.~\ref{eq:pot} is used for our study. The detailed nature of the
short-range potential is not important and we take it as a simple
square well potential. We concentrate in the range $0<\la<3/2$.

The von Neumann entropy ($S$) is defined as 
\begin{equation}
  S=-{\rm  Tr}\,(\rho\ln\rho)
\end{equation} 
where $\rho$ is the reduced density matrix for the ground state
$|\psi\rangle$ of a two particle system,
\begin{equation} 
  \rho(\textbf{r}_1,\textbf{r}_1')= {\rm Tr}_2\,\varrho(1,2)=\int \rmd^3
  \textbf{r}_2\ \langle\textbf{r}_1,\textbf{r}_2
  |\psi\rangle\langle \psi|\textbf{r}_1',\textbf{r}_2\rangle, 
  \label{eq:3} 
\end{equation} 
obtained from the two particle density matrix
\hbox{$\varrho(1,2)=\mid\!\psi\rangle\langle\!\psi\!\mid$} by
integrating out particle 2. The density matrix in $\ln\rho$ is to be
made dimensionless by appropriate powers of cut off $a$. 

As the interaction is translationally invariant, the reduced density
matrix $\rho(\textbf{r},\textbf{r}')\equiv
\rho(\textbf{r}-\textbf{r}')$ satisfies the eigenvalue
equation\cite{psmb},
\begin{equation}
  \int \rmd^3{\bf r}' \ \rho(\textbf{r}-\textbf{r}')\exp(-i{\bf q}\cdot {\bf
    r}')=\hat{\rho}({\bf q}) \exp(-i{\bf q}\cdot {\bf
    r}).
  \label{eq:eval} 
\end{equation}
The eigenfunction is $\exp(-i{\bf q\cdot r})$ with the eigenvalue
\begin{equation}
\hat{\rho}({\bf q}) =|\phi({\bf q})|^2,
\end{equation}
where we have put the center of mass wave vector to be zero without loss of
generality. Here $\phi({\bf q})$ is the normalized momentum-space wave
function of the relative co-ordinate.  Therefore, the von Neumann
entropy reads,
\begin{equation}
  S=-\int \rmd^3{\bf q}\ |\phi({\bf q})|^2\ \ln|\phi({\bf q})|^2.
\end{equation}

The reduced density matrix in the basis of momentum states $|{\bf
  k}\rangle$ has the form
\begin{equation}
  \rho=\int \rmd^3 {\bf k}\ |\phi({\bf k})|^2\ |{\bf k}\rangle\langle {\bf k}|\stackrel{\mathrm{def}}{=}
  \int \rmd^3{\bf k}\ \frac{e^{-\beta H_{ent}}}{Z}\ |{\bf k}\rangle\langle {\bf k}|,
\label{eq:2} 
\end{equation}
which makes the mixed state characteristic explicit.  Eq.~\ref{eq:2}
allows us to define $\rho$ as a thermal density matrix with an
entanglement Hamiltonian $H_{ent}$ at a fictitious inverse temperature
$\beta$ with $Z$ as the partition function. This thermal
correspondence makes the von Neumann entropy equivalent to the Gibbs
entropy of $H_{ent}$.

In Eq.~\ref{eq:2}, $H_{ent}$ is a {\it c}-number. Consider the
canonical partition function of a free particle at temperature $T$,
\hbox{$ Z\sim \int \rmd^dq \exp (-\beta H)\sim T^{d/2}, $} where
$H=\hbar^2q^2/2m$. Then the entropy becomes, $S=\ln Z\sim \ln T$,
which for very low temperature, $T\ra 0$ becomes negative. In another
way, one gets a constant specific heat $C$ from the equipartition
theorem, which then gives a logarithmic dependence on temperature of
the entropy, \hbox{$ S=\int^T (C/T) \rmd T\sim \ln T.$} The
Sackur-Tetrode constant, the dimensionless entropy of one mole of an
ideal gas at temperature $T=1K$ and one atmospheric pressure, is a
fundamental constant\cite{rmp}. Its value is $-1.164 8708$.  Note that
this fundamental entropy is negative.  Classical harmonic oscillator
is no exception. It is well-known that the condition $S\ge 0$ does not
hold for the classical continuous statistical mechanics\cite{lieb}.

\section{Analytical results}

In this section, we derive the asymptotic behaviour of $\phi(q)$. In
particular we find that the entropy is dominated by the outer part,
i.e. the excursion in the classically forbidden region, if the
unbinding transition is critical. This happens for $0<\la\le 1$. For
first order transition, the inner part also contributes significantly.

To find out the von Neumann entropy of two interacting particles in
the ground state, we first write the wave function as a product of
that for the center of mass (CM) and for the relative coordinate:
\begin{equation}
  \psi(\textbf{r}_1,\textbf{r}_2)=\Phi({\bf R})\,\varphi({\bf r}),
\label{eq:entdef}
\end{equation}
where $\Phi$ and $\varphi$ are the wave function in the CM
and relative coordinates respectively, defined by,
\begin{equation}
  \textbf{R}=\frac{m_1\textbf{r}_1+m_2\textbf{r}_2}
  {m_1+m_2} \quad {\rm and}\quad \textbf{r}=\textbf{r}_1-\textbf{r}_2.
\end{equation}
We proceed with the radial part putting the CM momentum as zero. The
CM is completely delocalized in space.

The ground state has zero angular momentum and therefore, only the
radial part of $\varphi({\bf r})$, $R(r)$, in the spherical polar
coordinate is needed. For this $s$-state, the radial part of the
Schrodinger equation then reads\cite{van}:
\begin{eqnarray}
  \frac{\pa^2 R}{\pa r^2}+\frac{2}{r}\frac{\pa R}{\pa r}+
  \left(V_0+E\right)R=0, \ \ {\rm for}\ \  r\!<\!a  \label{eq:Ri},\ \ \ \ \\
  {\rm and },\ \frac{\pa^2 R}{\pa r^2}+\frac{2}{r}\frac{\pa R}{\pa
    r}+\!\left(\frac{g}{r^2}+E\right)\!R=0, \ \ {\rm for}\ \ r\!>\!a\label{eq:Ro} ,\ \ \ \ 
\end{eqnarray} 
where $E$ is the ground state energy of the particle describing the
behaviour of the two particles in relative coordinate. The radial part of the wave
functions in the relative coordinate is then obtained by solving Eqs.
\ref{eq:Ri} and \ref{eq:Ro},
\begin{numcases}{ R(r) =}
  \frac{A}{r} \sin kr, \quad\quad\quad {\rm for}\ \ r\le a \label{eq:inR}\\
  \frac{B}{\sqrt{r}} H_\lambda^{(1)} (i\ka r), \ \ \ {\rm for}\ \ r\ge
  a ,\label{eq:ouR}
\end{numcases}
where $A$ and $B$ are the normalization constants, $H_\la^{(1)}$ is
the Hankel function of first kind, and  
\begin{equation}
  k^2=V_0-|E|,\ \ \ \ka^2=|E|.
\end{equation}
We choose $\lambda$ to be positive and it is given by
Eq.~\ref{eq:glam}. In the limit of $\ka\ra 0$, the unbinding
transition takes place. This makes our interest in studying the von
Neumann entropy in this limit.

At $r=a$, the continuity of the wave functions gives,
\begin{equation} 
  \frac{A}{a} \sin ka=\frac{B}{\sqrt{a}}
  H_\lambda^{(1)} (i\ka a),  \label{eq:1}
\end{equation} 
and the matching condition of the derivative of the wave function
gives,
\begin{eqnarray} 
  ak\cot ak = i\ka a
  \frac{ H_{\lambda-1}^{(1)} (i\ka a)}{ H_\lambda^{(1)} (i\ka a)}
  -\lambda + \frac{1}{2}\ ,
\label{eq:kc}
\end{eqnarray} 
which determines the value of $k$ for a given $\ka$. Given the values
of $\lambda$ and $a$, one can get the threshold or minimum value of
$k$, $k_m$, for just one bound state. For $\ka= 0$,
\begin{equation}
  a k_c \cot ak_c=\frac{1}{2}-\la,
\label{eq:4}
\end{equation}
is the condition for the transition point when the ground state energy
$E\ra 0$.  Eq.~\ref{eq:4} has always a solution for $\lambda\ge 0$.

Now consider a small deviation from the critical value of $k$,
$k=k_c-\delta$ where $\delta\sim V_0-V_c$. Then, from Eq.~\ref{eq:kc},
\begin{equation}{
  (a k_c-a\delta)\cot (a k_c-a\delta) \sim 
}\cases{
  (\kappa a)^{2\lambda},\, \quad {\rm for}\ \ \lambda<1,\\
  (\kappa a)^2, \quad\quad {\rm for}\ \ \lambda>1,}
\end{equation} 
or,
\begin{equation}{
  |E|=\kappa^2 \sim
}\cases{
  \delta^{1/\lambda}, \quad\quad \quad\quad\quad {\rm for}\ \ 0<\lambda<1,\\
  \delta + O(\delta^{1/(\la-1)}), \quad \quad {\rm for}\ \ \lambda>1.}
\end{equation} 
These show that as $V_0\rightarrow V_c\equiv k_c^2$, $E$ remains
continuous, as it should. For $\lambda<1$, $E$ approaches zero
tangentially, while for $\lambda>1$, there is a nonzero slope at
$\ka=0$. This discontinuity of slope classifies the $\la>1$ transition
as first order. Despite that, the higher derivatives on the bound side
$\partial^n E/\partial \delta^n$ would show divergences like a
critical point.

The normalization constants $A$ and $B$ are found by using the
continuity condition and taking the limit $\ka\ra 0$ (see Appendix for
details),
\begin{equation}{
  |A|^2 \sim}\cases{
    (a\ka)^{2-2\la}/a, \ \ {\rm for}\ \ \lambda< 1,\\
    1/a, \quad\quad\quad\quad {\rm for}\ \ \lambda> 1,     }   
\end{equation} 
and
\begin{equation}{
  |B|^2 \sim
}\cases{
  \ka^2, \quad\quad\quad\quad\quad {\rm for}\ \ \lambda< 1,\\
  \ka^2 (a \ka)^{2\la-2}, \ \ \, \ {\rm for}\ \ \lambda> 1 .}
\end{equation} 
At $\la=1$, there are log corrections which we do not get into here.
The log correction appears in the Necklace model for polymers whenever
the reunion exponent $\Psi$ (Eq.~\ref{eq:glam}) is an integer. The
log appears in Eq.~\ref{eq:kc} via $H_0^{(1)}$ for $\la=1$.  Now one
knows the full wave function and its limiting $\ka$ behaviour.

The reduced density matrix has eigenvalues $|\phi({\bf q})|^2$, where
${\bf q}$ is
the momentum space variable. To get these eigenvalues, the Fourier
transformation of the wave-function needs to be done,
\begin{eqnarray} 
  \phi(q) =
  \frac{1}{(2\pi)^{3/2}}  \int \rmd^3 r\ e^{i{\bf q}.{\bf r}}R(r) 
  = \phi_{\rm i}(q) + \phi_{\rm o}(q) ,\ \ \ \ 
\end{eqnarray} 
where the subscript ${\rm i,o}$ refer to the inner ($r<a$) and the outer
($r>a$) part. The Fourier transform of the inner part (Eq.~\ref{eq:inR}) is
\begin{equation} 
  \phi_{\rm i}(q) =
  \frac{A}{q}\frac{1}{\sqrt{2\pi}}\left[\frac{\sin(k-q)a}{k-q}-
    \frac{\sin(q+k)a}{q+k}\right], \label{eq:wfi} 
\end{equation} 
and of the outer part (Eq.~\ref{eq:ouR}) is
\begin{eqnarray} 
  \hskip -1cm
  \phi_{\rm o}(q) 
  &=& |B|\ \kappa^{-5/2}\ \frac{\sqrt{2}}{\pi}\ \Gamma
  \left(\frac{5}{4}+\frac{\lambda}{2}\right)
  \Gamma\!\!\left(\frac{5}{4}-
    \frac{\lambda}{2}\right){}_2F_1\left(\frac{5}{4}+
    \frac{\lambda}{2}, \frac{5}{4}-\frac{\lambda}{2};
    \frac{3}{2};-\frac{q^2}{\ka^2}\right)\no\\
  &&-|B| \int_0^a \rmd r \sqrt{r} \frac{\sin qr}{q} K_\lambda (\ka
  r),  \label{eq:wfo}
\end{eqnarray} 
where $_2F_1$ is the hypergeometric function and $K_\la$ is the
modified Bessel function. The last integral in Eq.~\ref{eq:wfo} is
convergent for all $\la<1$ and therefore can be ignored in the
$\ka\ra 0$ limit.

The limiting small $\ka$ dependence of the inner and the outer parts of the
wave function from Eqs. \ref{eq:wfi} and \ref{eq:wfo} are of the form
\begin{equation}{ 
  \phi_{\rm i}(q)
  = }\cases{
    {(a\ka)}^{1-\la}a^{3/2}f_{\rm i}(aq),
    \ \ \ \ \ \, {\rm if\ \ \la< 1},\\
    a^{3/2}f_{\rm i}(aq),
    \quad\quad\quad\quad{\rm if\ \ \la> 1} ,    }
      \label{eq:insc}
\end{equation}
and
\begin{equation}{
 \phi_{\rm o}(q) = 
}\cases{
  \ka^{-\frac{3}{2}}\, f_\la(q/\ka),\ \quad\quad\quad
  \ \, {\rm if\ \ \la< 1},\\
  \ka^{-\frac{3}{2}}\,(a\ka)^{\la-1} f_\la(q/\ka),\ \ \ \ \ {\rm
    if\ \ \la> 1} ,}
\label{eq:ousc}
\end{equation} 
where $f_{\rm i}$ and $f_\la$ are well-behaved functions.
Eq.~\ref{eq:insc} is for large $q$.

From Eqs.~\ref{eq:insc} and \ref{eq:ousc}, we see that the double
limit $\ka\ra 0$, $\la\ra 1$ is singular because of the term
$(a\ka)^{1-\la}$. The dependence of entropy on the order of the limits
is discussed later on. This identifies $(\ka=0,\la=1)$ as a special
point. From this we also identify $(1-\la)\ln (a\ka)$ as an
appropriate scaling variable. This scaling variable will occur below
in the analysis of the numerical results.

For $\la<1$, i.e., $1-\la>0$, $(a\ka)^{1-\la}\rightarrow 0$ as
$\ka\rightarrow 0$ and therefore, the contribution of outer part
dominates over the inner part in the von Neumann entropy.  Without
much loss, one can then write the entropy with the outer part only
(Eq.~\ref{eq:ousc}),
\begin{eqnarray}
  S&\approx&-\int \rmd^3 q\ |\phi|^2\ln|\phi_{\rm o}|^2 \no\\
  &=& 3\ln a\ka+ c_\la,\quad ({\rm for}\quad \la<1)
\label{eq:ee}
\end{eqnarray}
with 
\begin{equation}
  c_\la= \int \rmd x\, x^2 f_\la(x)\ln f_\la(x).
\end{equation} 
We introduced $a$ in Eq.~\ref{eq:ee} to make the argument of $\ln$
dimensionless, mentioned earlier. As per our interest, we extract the
$\ka$-dependent term and call the rest $c_\la$, which is a function of
other parameters. The main result is that there is a log divergence of
$S$ as $\ka\ra 0$.

\section{Exact numerical results}

To study the nature of the entanglement entropy, over the whole range
of $\la$ we take recourse to exact numerical calculation using {\sc
  mathematica} for the $3$-dimensional potential well. We cross-check
our prediction of Eq.~\ref{eq:ee} and then show a multicritical
scaling that covers the range $0<\la<3/2$.

\subsection{Protocol}
Although $V_0$ is the tuning parameter, it is more convenient to use
the length scale as the independent parameter. With this treading of
$\ka$ for $V_0$, our protocol is like this: Given the values of $\ka$
and $\la$, the value of $k_m$ was determined from Eq.\ref{eq:kc}, with
$k_m<\pi$ that assures us the ground state. As $\ka\ra 0$, $k_m\ra
k_c$. Then the corresponding normalization coefficients $A$ and $B$
were found by using the normalization condition and the continuity
equation, i.e., by doing the $r$-integrations of the inner and the
outer parts of the wave function in Eq.~\ref{eq:B}. These constants are
used in the Fourier transformed inner and outer parts of the wave
function, Eqs.  \ref{eq:wfi} and \ref{eq:wfo}, to calculate the von
Neumann entropy.  In the final integration for $S=-{\rm Tr}\ \rho
\ln\rho$, we put an upper cut off making sure that the final numbers
are independent of this choice of cut off. Also the intervals of the
integration range have been chosen carefully especially for
$q\sim\ka$. This gives numerically exact numbers for the entropy for
the given $\ka$ and $\la$. This procedure is repeated for various
$\la$ and $\ka$. We set $a=1$.

\sla

\subsection{Behaviour of the von Neumann entropy $S$}

\subsubsection{$\la$ dependence:}
The plots of the numerical values of the von Neumann entropy $S$
against $\ln\ka$ and $\la$ show different behaviours of entropy in
different ranges of $\la$, {\it viz.}, $\la<1$, $\la>1$ and $\la=1$.

Let us first look at the plot of $S$ {\it vs.} $\la$ in fig.
\ref{fig:sla}, where different lines represent different
values of $\ka$.  For $\la<1$, the von Neumann entropy for small $\ka$
saturates to a negative value as $\la$ is varied and that saturation
value depends on the value of $\ka$. The smaller the value of $\ka$,
the more negative is the entropy, and $\ka\rightarrow 0$ takes the
saturation value to negative infinity. The long-range part of the
potential is attractive for $\la>0.5$ and repulsive otherwise.  But
the entropy shows no signature as it crosses $\la=0.5$. On the other
hand, for $\lambda>1$ where the transition becomes first order, the
entropy does not decrease much with $\ka$, rather becomes independent
of $\ka$. It remains finite for $\la>1$ and diverges at $\la=1$ like
the black dashed curve in Fig. \ref{fig:sla}. $S$ becomes
positive at $\la\sim 1.3$.  It seems that this point has no
significance otherwise.

\skap

\subsubsection{$\ka$ dependence:}
The behaviour of the von Neumann entropy with $\la$ and $\ka$ becomes
more clear when one looks at the plot of $S$ {\it vs.}  $\ka$ (Fig.
\ref{fig:skap}). This plot shows the different characteristic
behaviours of $S$ in the three distinct ranges of $\la$: $\la<1$,
$\la=1$ and $\la>1$. For small $\ka$, all $\la<1$ curves have slope
$3$ when plotted against $\ln\ka$, i.e. for $\la<1$, the entropy is of
the expected form $3\ln\ka+c_\la$ which is shown from analytical
calculations. To get $3\ln\ka$, one has to see below some value of
$\ka$, and as $\la$ approaches one, even smaller $\ka$ needs to be
considered.  But no matter how close to $1$ is the value of $\la$, one
gets $3\ln\ka$ until $\la<1$. Exactly at $\la=1$, the slope changes
suddenly to $3/2$ and hence
\begin{equation}
  S =\frac{3}{2} \ln\ka +c_1,\quad {\rm for}\quad \la=1.
\end{equation}

A somewhat different behaviour is seen for the rest with $\la>1$ (inset
of Fig.  \ref{fig:skap}). For small $\ka$, the curves reach a
$\la$-dependent constant value and do not change with $\ka$.  Clearly
the entropy has no $\ka$-dependence for $\la>1$ and it is finite.  By
definition, these constant values are $c_{\la}$ and $S
(\la>1)=c_{\la}$.  So, we see that there are three classes: 
\begin{equation}{
  S =
}\cases{
  3\ln\ka+c_\la \quad {\rm for}\ \la<1,\\
  \frac{3}{2}\ln\ka +c_1 \!  \quad {\rm for}\ \la=1,\\
  c_\la \quad \ \quad \quad \quad {\rm for}\ \la>1.}
\label{eq:scla}
\end{equation}

\subsubsection{On $c_\la$:}

\clam 

Now we have knowledge of the $\ka$ dependent part in the von Neumann
entropy for different $\la$.  The next question is how the $c_\la$
behaves with $\la$, and if they have different nature in different
regimes of $\la$. So, we collect the $c_\la$s according to
Eq.~\ref{eq:scla}, except for $\la=1$, and plot against $\la$.  This plot
(Fig.~\ref{fig:clam}) shows a divergence at $\la=1$ indicating
that $(1-\la)$ is an important quantity. The data points are fit into
the function
\begin{equation}
  c_\la=m+n/(1-\la),
\end{equation} 
via $m$ and $n$, and the fitted set of parameters are $(4.52,1.38)$
and $(3.76,1.48)$ for $\la$ greater and less than one respectively.
The divergence of $c_\la$ at $\la=1$ leads to the possibility of a
reduction of the slope of $S$ from $3$ to $3/2$ when plotted against
$\ln\ka$.

\col

\subsubsection{Data collapse:}
We noted that for $\la>1$, $c_\la$ and hence the entropy itself, has a
$(1-\la)$ dependence and for $\la<1$, the entropy has a $\ln\ka$ term
with $c_\la=f(1-\la)$. It was pointed out in Sec 3, below
Eq.~\ref{eq:ousc} that $(1-\lambda) \ln \kappa$ seems to be a scaling
variable. We therefore look at the plot of the entropy {\it vs.}
$(1-\la)\ln\ka$. The entropy has different behaviours on the two sides
of the $\la=1$ making it a special point. Also, it has a separate
scaling behaviour. This drives us to plot
$(S_\la-S_1)/(\frac{3}{2}\ln\ka)$ against $(1-\la)\ln\ka$. We see a
good data collapse (see Fig.~\ref{fig:col}) for various sets of data
of Fig~\ref{fig:sla}. Hence, one can write the scaling form of
von Neumann entropy:
\begin{equation}
  (S_\la-S_1)/\frac{3}{2}\ln\ka={\cal F}((1-\la)\ln\ka).
\label{eq:scaling}
\end{equation}
Fig.~\ref{fig:col} shows that $(S_\la-S_1)/\frac{3}{2}\ln\ka$ reaches
$+1$ for small enough $\ka$ for $\la<1$ and $-1$ for $\la>1$.  Once we
get the scaling behaviour of the entropy at $\la=1$, the same away from
this special point can also be obtained.

\section{Discussion and conclusion}
In this paper we studied the von Neumann entropy $S$, the most common
measure of the entanglement entropy, for an inverse square potential
in three dimensions. The entropy behaves in different ways for three
different ranges of modified interaction strength $\la$ and given by
Eq.~\ref{eq:scla}.  For $\la<1$ and $\la=1$, the entropy has a
diverging nature as one approaches quantum critical point by tuning
$\ka$.  The behaviour of entropy is completely different for
$\la>1$, where the $\ka$-dependence of the entropy vanishes. There is
a $\frac{1}{1-\la}$ divergence in the entropy. These three distinct
classes collapse onto a single curve when
$(S_\la-S_1)/\frac{3}{2}\ln\ka$ is plotted against $(1-\la)\ln\ka$.
This data collapse indicates that there is a common scaling behaviour
of the entropy for any $\la$ and that $\la=1$ is special. Because of
the diverging factor dependent on ($1-\la$), one has to be careful in
taking required limit of $\ka\ra 0$ as that would give a log
correction in entropy for $\la=1$. For $g>1/4$, $\la$ is imaginary
which we do not consider here. In this paper we focused on the
multicritical point at $\lambda=1$.  There is one more multicritical
point at $\lambda=0$ with KT transition. This case will be discussed
elsewhere. It would be interesting to study the entanglement behaviour
for a discretized version of the model.

\limit

The nature of the divergence of the entanglement entropy at $\la=1$
depends on the path of approaching $\la=1$ in a $\la$-$\ka$ plane.
Diagrammatically it has been shown in Fig.~\ref{fig:limit}. If we take
the limit $\ka\ra 0$ first and then $\la=1$, the entropy diverges like
$1/(1-\la)$ (shown by red line (a) in Fig.~\ref{fig:limit}), and like
$\ln\ka$ for the other way around (see the blue line (b) in the same
figure). For $\la<1$, the $\ka\ra 0$ line corresponds to $S=-\infty$,
but for $\la>1$ the same line gives a finite value for entropy.  The
path dependence of Fig.~\ref{fig:limit} summarizes the features of the
entanglement entropy, with $\lambda=1,\kappa=0$ as a special point
controlling the behaviour in its neighborhood. The data collapse of
Fig.~\ref{fig:col}, then, suggests that the paths should be classified
by the constant value of $X=(1-\lambda) \ln\kappa$.

For $\la<1$, restricting to the critical case, we see $\rho(q)\sim
|\phi(q)|^2$. These are the eigenvalues of the density matrix. Now the
reduced density matrix $\rho$ describes a mixed state, though the full
ground state is pure. Being a mixed state, we may represent $\rho$ as
a ``thermal'' density matrix, $\rho\sim\exp(-\beta H_{ent})$, as done in
Eq.~\ref{eq:2}. Since the entanglement spectrum is known, we have 
\begin{equation}
  \beta H_{ent} \approx \ln |_2F_1|^2\approx \frac{1}{2}
\frac{q^2}{\ka^2},\quad
  {\rm for}\ q\ra 0 ,
\end{equation}
identifying $\beta=1/\ka^2$ and $H_{ent}=q^2/2$. As mentioned before,
for this $H_{ent}$ the Gibbs entropy is \hbox{$\sim\frac{d}{2}\ln T$}.
Since in this case $T\simeq \ka^2$, we find the von Neumann entropy
$S\sim d\ln\ka$.

\appendix

\section{Calculation of the normalization constants $A$ and $B$}
\label{sec:calc-norml-const}

The normalization constants $A$ and $B$ are found by using the continuity
condition and taking limit $\ka\ra 0$ in the normalization condition,
\begin{equation} 
  4\pi\left[\int_0^a |A|^2
    \sin^2 kr\, \rmd r +
    \int_a^\infty r |B {H_\lambda^{(1)}} (i\ka r)|^2\, \rmd r\right]=1. 
\end{equation} 
Replacing $A$ by $B$, by using the continuity equation,
Eq.~(\ref{eq:1}), we get,
\begin{equation} 
\hskip -1cm  \left[\left(2\pi
      a-\frac{\pi}{k}\sin{2ak}\right)\frac{a |H_\la^{(1)}(i\ka
      a)|^2}{\sin^2 ka}\ + 4\pi \int_a^\infty r  |H_\lambda^{(1)}
    (i\ka r)|^2\, \rmd r\right]|B|^2= 1 .
\label{eq:B}
\end{equation} 
Now we use the form of the Hankel function in the limit
$\ka\rightarrow 0$,
\begin{equation}
\mid H_\la^{(1)}(i\ka
      r)\mid^2 \sim \frac{2^\la\Gamma^2(\la)}{\pi^2}\,(\ka r)^{-2\la},
\label{eq:Hlim}
\end{equation}
and rewrite the outer part integral in the normalization condition in
a simpler form, {\it viz.},
\begin{eqnarray}
  && \hskip -2cm\int_{\ka a}^\infty r \mid H_\lambda^{(1)}
   (i r)\mid^2\, \rmd r \no\\
  && \hskip -2.5cm = \int_{\ka a}^1 \left[| H_\lambda^{(1)}
     (i r)|^2-\frac{2^\la\Gamma^2(\la)}{\pi^2}\,r^{-2\la}\right]\, 
   \rmd r
   + \int_{\ka a}^1 \frac{2^\la\Gamma^2(\la)}{\pi^2}\,r^{-2\la}\,
   \rmd r + \int_{1}^\infty r | H_\lambda^{(1)}(i r)|^2\, \rmd r\no\\
   && \hskip -2.5cm = \frac{\Gamma^2(\la)}{\pi^2\,2^{1-\la}} 
   \frac{1-(a\ka)^{2(1-\la)}}{1-\la}  + ...\ \  .
\label{eq:oint}
\end{eqnarray}
Putting Eqs. (\ref{eq:oint}) and (\ref{eq:Hlim}) in Eq. (\ref{eq:B})
one gets,
\begin{equation}
\hskip -1cm   |B|^2=\frac{\pi}{2^{\la+1}\Gamma^2(\la)}
  \left[
    \left( 
      \frac{\la^2+ a^2k^2-1/4}{k^2}
    \right)(a\ka)^{-2\la}
    + \frac{1}{\ka^2}
    \frac{1-(a\ka)^{2(1-\la)}}{(1-\la)} \right]^{-1},
\label{eq:BB}
\end{equation} 
which in the extreme limit of $\ka\rightarrow 0$ gives the $\ka$
dependence of $B$,
\begin{equation}{
  |B|^2 \sim
}\cases{
  \ka^2, \quad\quad\quad\quad\quad {\rm for}\ \ \lambda< 1,\\
  \ka^2 (a \ka)^{2\la-2}, \ \ \, \ {\rm for}\ \ \lambda> 1 .}
\end{equation} 
Once $B$ is obtained, the $\ka$-dependence of the other constants $A$
can be found by using Eq.~(\ref{eq:1}) as,
\begin{equation}{
  |A|^2 \sim}\cases{
    (a \ka)^{2-2\la}/a, \ \ {\rm for}\ \ \lambda< 1,\\
    1/a, \quad\quad\quad\quad {\rm for}\ \ \lambda> 1.    }    
\end{equation}

\end{document}